\documentclass[aps,prl,twocolumn,tightenlines,amsmath,amssymb,superscriptaddress]{revtex4-1}

\usepackage{graphicx}
\usepackage{amsmath}
\usepackage{verbatim}
\usepackage{subfigure}
\usepackage{epstopdf}
\usepackage{color}
\usepackage{natbib}
\usepackage[normalem]{ulem}

\definecolor{mygrey}{gray}{0.35}
\definecolor{myblue}{rgb}{0.2,0.2,0.8}
\definecolor{myzard}{cmyk}{0,0,0.05,0}
\definecolor{mywhite}{rgb}{1,1,1}
\definecolor{myred}{rgb}{1,0.,0.3}
\definecolor{Burgundy}{rgb}{0.5,0.,0.13}

\newcommand{\be}{\begin{equation}}
\newcommand{\ee}{\end{equation}}

\newcommand{\mred}[1]{{\color{myred}#1}}

\usepackage[colorlinks=true,citecolor=myblue,linkcolor=myred,urlcolor=myblue]{hyperref}

\begin{document}

\title{Sequential generation of linear cluster states from a single photon emitter
}
\author{D. Istrati}
\altaffiliation[]{These authors contributed equally to this work.}
\affiliation{Racah Institute of Physics, Hebrew University of Jerusalem,
Jerusalem 91904, Israel}
\author{Y. Pilnyak}
\altaffiliation[]{These authors contributed equally to this work.}
\affiliation{Racah Institute of Physics, Hebrew University of Jerusalem,
Jerusalem 91904, Israel}
\author{J. C. Loredo}
\affiliation{CNRS Centre for Nanoscience and Nanotechnology, Universit\'{e} Paris-Sud,
Universit\'{e} Paris-Saclay, Palaiseau, France
}
\author{C. Ant\'{o}n}
\affiliation{CNRS Centre for Nanoscience and Nanotechnology, Universit\'{e} Paris-Sud,
Universit\'{e} Paris-Saclay, Palaiseau, France
}
\author{N. Somaschi}
\affiliation{Quandela, Palaiseau, France
}
\author{P. Hilaire}
\affiliation{CNRS Centre for Nanoscience and Nanotechnology, Universit\'{e} Paris-Sud,
Universit\'{e} Paris-Saclay, Palaiseau, France
}
\affiliation{Universit\'{e} Paris Diderot, Paris, France
}
\author{H. Ollivier}
\affiliation{CNRS Centre for Nanoscience and Nanotechnology, Universit\'{e} Paris-Sud,
Universit\'{e} Paris-Saclay, Palaiseau, France
}
\author{M. Esmann}
\affiliation{CNRS Centre for Nanoscience and Nanotechnology, Universit\'{e} Paris-Sud,
Universit\'{e} Paris-Saclay, Palaiseau, France
}
\author{L. Cohen}
\affiliation{Racah Institute of Physics, Hebrew University of Jerusalem,
Jerusalem 91904, Israel}
\author{L. Vidro}
\affiliation{Racah Institute of Physics, Hebrew University of Jerusalem,
Jerusalem 91904, Israel}
\author{C. Millet}
\affiliation{CNRS Centre for Nanoscience and Nanotechnology, Universit\'{e} Paris-Sud,
Universit\'{e} Paris-Saclay, Palaiseau, France
}
\author{A. Lema\^{i}tre}
\affiliation{CNRS Centre for Nanoscience and Nanotechnology, Universit\'{e} Paris-Sud,
Universit\'{e} Paris-Saclay, Palaiseau, France
}
\author{I. Sagnes}
\affiliation{CNRS Centre for Nanoscience and Nanotechnology, Universit\'{e} Paris-Sud,
Universit\'{e} Paris-Saclay, Palaiseau, France
}
\author{A. Harouri}
\affiliation{CNRS Centre for Nanoscience and Nanotechnology, Universit\'{e} Paris-Sud,
Universit\'{e} Paris-Saclay, Palaiseau, France
}
\author{L. Lanco}
\affiliation{CNRS Centre for Nanoscience and Nanotechnology, Universit\'{e} Paris-Sud,
Universit\'{e} Paris-Saclay, Palaiseau, France
}
\affiliation{Universit\'{e} Paris Diderot, Paris, France
}
\author{P. Senellart}
\affiliation{CNRS Centre for Nanoscience and Nanotechnology, Universit\'{e} Paris-Sud,
Universit\'{e} Paris-Saclay, Palaiseau, France
}
\author{H. S. Eisenberg}
\affiliation{Racah Institute of Physics, Hebrew University of Jerusalem,
Jerusalem 91904, Israel}


\begin{abstract}
Light states composed of multiple entangled photons---such as cluster states---are essential for developing and scaling-up quantum computing networks~\cite{Raussendorf01,Rudolph17,Zwerger12,Zwerger16,Azuma15}. Photonic cluster states with discrete variables can be obtained from single-photon sources and entangling gates, but so far this has only been done with probabilistic sources constrained to intrinsically-low efficiencies~\cite{Walther05,Pilnyak17,Zhong18}, and an increasing hardware overhead. Here, we report the resource-efficient generation of polarization-encoded, individually-addressable, photons in linear cluster states occupying a single spatial mode. We employ a single entangling-gate in a fiber loop configuration~\cite{Pilnyak17} to sequentially entangle an ever-growing stream of photons originating from the currently most efficient single-photon source technology---a semiconductor quantum dot~\cite{Senellart17,Somaschi16,Ding16,Loredo16,Wang16}. With this apparatus, we demonstrate the generation of linear cluster states up to four photons in a single-mode fiber. The reported architecture can be programmed to generate linear-cluster states of any number of photons with record scaling ratios, potentially enabling practical implementation of photonic quantum computing schemes.
\end{abstract}

\maketitle

Optical quantum technologies include a wide range of applications, from quantum sensing~\cite{Giovannetti11,Pittman95}, to quantum communication~\cite{BB84} and computing~\cite{Knill01,Raussendorf01,Walther05}. Entanglement is the most common resource for these applications~\cite{EPR35}, exploiting  various degrees-of-freedom, e.g. polarization, time-frequency, orbital angular momentum, and spatial modes. The generation of high quality large cluster states can be used for photonic one-way quantum computing~\cite{Raussendorf01, Walther05,Rudolph17}, which is favourable compared to the widely used Turing-like gate based model in solid-state systems~\cite{Knill01}. Moreover, photonic cluster states have been proposed to implement measurement-based quantum communication networks~\cite{Zwerger12,Zwerger16,Azuma15}, an architecture that promises long distance quantum communication at higher rates compared to other memory-based counterparts. 

\begin{figure}[t]
\includegraphics[width=1\columnwidth]{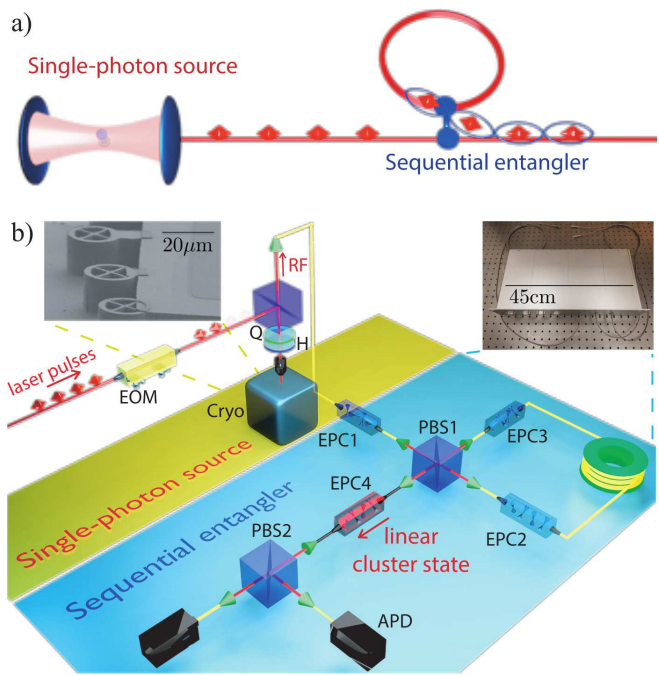}
\caption{\label{Fig1} {\textbf{A fibered source of linear photonic cluster states} \textbf{(a)} Working principle of the source: single-photons generated in successive time bins are sent into an apparatus where a delay-loop stores a photon until it meets with the next one at the entangling gate. \textbf{(b)} Top {inserts}: the physical implementation. A single InGaAs quantum dot photon source in an electrically-connected cavity, and an all-fibered entangling optical circuit in a 19'' box. Bottom: Detailed experimental setup - see text for details.}\label{fig1}}
\end{figure}

Cluster states are a special class of multi-partite graph states that show improved robustness to loss compared to GHZ-, or W-states~\cite{Dur04,Hein05,GHZ90}. They can be generated using single-photon sources and entangling operations. So far, they have been implemented using probabilistic nonlinear sources of photon pairs up to six photons, by the $\chi^{(2)}$ parametric down-conversion (PDC), and the $\chi^{(3)}$ four-wave mixing processes, or, for  continuous variable encoding,  by multi-mode squeezing in optical parametric oscillators~\cite{Walther05,Lu2006,Bell14,Reimer19,Larsen19,Asavanant19}. Photonic GHZ states of up to twelve entangled photons have also been obtained in this manner~\cite{Zhong18}. The common approach consists of multiple sources and multiple well balanced paths to manipulate the photons. In this approach, it is important to operate in a regime where the probability to generate a pair is low (typically few percents) in order to limit multi-pair emission. These low source efficiencies make the protocols difficult to scale up to large photon numbers, in addition to an increased resource budget when employing multiplexed schemes~\cite{Mendoza16,Kaneda19}. A more scalable way to produce large photonic cluster states has been proposed in 2009~\cite{Lindner09}, making use of a single quantum emitter embedding a spin acting as a quantum memory. A proof-of-concept experimental demonstration has been provided with a semiconductor quantum dot up to 2 photons~\cite{Schwartz16}. However, this was obtained at very low generation rates, and with considerable challenges to allow further scalability, such as the need for longer spin coherence times and efficient polarization-independent photon extraction.

In this work, we demonstrate an approach for the generation of photonic cluster states that takes advantage of a newly available technology for single-photon generation---semiconductor quantum dots (QDs)~\cite{Senellart17}---and a recent proposal for entanglement generation based on temporal delay-loops~\cite{Pilnyak17}. Quantum dots generate single-photons on demand with near-unity indistinguishability~\cite{He13,Somaschi16,Ding16,Snijders18}, and high single-photon purity. In addition, they can have high in-fiber brightness (defined as the probability to have a  single-photon coupled into a single-mode fiber per excitation pulse), typically one order-of-magnitude larger than heralded single-photon sources~\cite{Somaschi16,Ding16}. This allows for an exponential increase in multi-photon generation rates, which has already been used for Boson sampling~\cite{Loredo17,Wang17,Wang19b}, and on-chip quantum walks~\cite{Anton19}. We employ a fiber delay-loop apparatus to sequentially entangle photons successively generated by a bright QD single-photon source. Linear photonic cluster states of two, three and four photons are obtained with high rates. Our compact entangling apparatus allows for both entanglement generation and polarization state analysis. Our experimental demonstration brings the record for the number of entangled photons from a single emitter from two photons~\cite{Santori04,He13, Gazzano13} to four photons with a fourfold generation rate of $\sim$10~Hz. Additionally, we define a parameter, the scaling-ratio, to quantify prospects of scalability, and to allow comparison between different implementations.

Figure~\ref{fig1}\mred{a} presents the principle of the proposed scheme. A single quantum dot positioned in an optical cavity serves as an efficient single-photon source. Periodical excitation with optical pulses leads to the emission of a stream of single-photons. The separated emission times enable individual addressability of each photon. The single-photons are then sent into an entangling gate, where the time between emissions is tuned to match the length of a delay-loop, that serves as a quantum memory. As a result, the combined system constitutes a source of linear cluster states encoded in the polarization degree-of-freedom, and individually addressable in the time domain. With this protocol, linear cluster states of any length can be produced, controlled by the number of consecutive photons sent into the entangling apparatus. Figure~\ref{fig1}\mred{a} depicts the case for 4 photons. All entangling operations occur at the same entangling gate, hence the low resource requirements of our approach.

Figure~\ref{fig1}\mred{b} depicts the physical implementation of our protocol. The single-photon sources used here are made of a single InGaAs quantum dot deterministically positioned in an electrically connected pillar cavity~\cite{Nowak14,Somaschi16} with optical resonances around 925$\,$nm. The emitter-cavity coupling ensures efficient collection of photons through accelerated spontaneous emission into the cavity mode. The QD transition is coherently controlled with resonant excitation pulses from a Ti:Sapphire laser operating at 81~MHz repetition rate. Maximum source efficiency is achieved by setting the excitation at $\pi$-pulse level, and the single photons are collected in a crossed polarization scheme. In this work, we used several sources with various characteristics: sources based on either neutral or charged excitons, with in-fiber brightness between $4\%$ and $15\%$. The detected single-photon rate out-of-fiber varies from $0.8$ to $3$~MHz using standard silicon avalanche photon detectors (APD) with 25\% detection efficiency typically. The degree of indistinguishability $M$ for photons generated by a single source at various time delays was measured for all sources, and varies from $0.77$ to $0.95$, the higher values being obtained with spectral filtering to remove contribution from phonon sidebands (see Supplementary Information). The laser pump driving the single-photon source is modulated by an electro-optic intensity modulator (EOM) that creates the required sequence for entangling the desired number of photons.

The entangling apparatus is implemented in an all-fibered compact device, packaged into a standard 19" rack mountable box, see inset in Fig.~\ref{fig1}\mred{b}. It has one single-mode fiber input, a fiber delay-loop about 15~m long, and one ``fusion" gate~\cite{Browne05} implemented by a polarizing beam-splitter (PBS). The output of this first PBS1 is sent to a single-mode fiber where the entangled photons emerge from. The analysis setup is also included in the same box. It consists of another projecting PBS2, whose two fiber outputs exit the box. Four electrically-driven polarization controllers, labelled EPC${i}$, contain four voltage-controlled birefringent elements each. They align the photon polarization between the fiber segments. EPC1 is positioned before the delay loop, EPC2 and EPC3 inside the loop, and EPC4 after the loop as part of the analysis setup. At the output of PBS2, two single-photon detectors are temporally synchronised to the laser clock frequency, and the time analysis of the state is controlled using a custom-designed FPGA controller.

\begin{figure}[t]
		\includegraphics[width=0.5\textwidth]{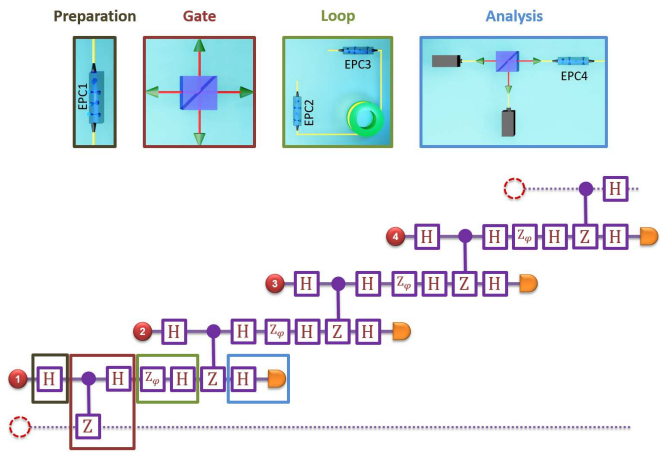}\vspace{-3mm}		
		\caption{\label{Fig2}Quantum circuit representation of the entangling scheme. The table maps the used logical operations, the Hadamard $\mathbf{H}$ and $\mathbf{Z_\varphi}$ transforms and the controlled-phase gate with their corresponding physical elements from Fig.~\ref{fig1}. First, each temporally separated photon (red disks) undergoes a state preparation step at EPC1 (black box) before entering the entangling gate (red box) and the delay loop with EPC2 and EPC3 (green box). When exiting the source, the photons reach the analysis step (blue box) of EPC4 and the photon polarization sensitive detection. Thus, the independent four successive incoming photons are transformed into a four-photon cluster state. The dashed circles/lines depict the absent photons right before and after the injected photon sequence.}
	\vspace{-4mm}	
	\label{fig2}
	\end{figure} 

The sequential operation of the entangling apparatus can be described in the following manner. The excitation laser is switched on and off using the EOM to create a series of consecutive laser pulses, corresponding to the temporal sequence of single-photons to be entangled. Each modulated pump sequence is preceded by two empty cycles in order to ensure that the delay loop is empty at the beginning of the protocol. The generated single-photons are injected into the sequential entangler, see Fig.~\ref{Fig1}\mred{b}, and their state is set to the diagonal polarization relative to PBS1 orientation $|p\rangle{=}\frac{1}{\sqrt{2}}(|h\rangle{+}|v\rangle)$ via EPC1, where $|h\rangle$ and $|v\rangle$ designate the horizontal and vertical polarization states, respectively. Similarly, the anti-diagonal state is defined as $|m\rangle{=}\frac{1}{\sqrt{2}}(|h\rangle{-}|v\rangle)$. The first photon has a 50\% chance of entering the delay loop through PBS1. The protocol succeeds if this first photon is transmitted through PBS1. Inside the loop, EPC2 controls a birefringent phase $\varphi$, and EPC3 rotates the $|h\rangle$ state of EPC2 orientation to the $|p\rangle$ state of PBS1. Each photon is delayed for $\tau{\simeq}74$~ns, corresponding to 6 laser cycles. When the photon inside the loop arrives to PBS1, it is timed to entangle with a new photon from the single photon source. This is achieved by fine adjustment of the pump laser repetition rate.

Post-selecting each photon to exit from a different port of the entangling PBS1, the two diagonal photons are projected onto the maximally entangled state
\be\label{phi}|p\rangle\otimes|p\rangle\xrightarrow[selection]{post}|\phi^+\rangle =\frac{1}{\sqrt{2}} \left(|h_1h_2\rangle+|v_1v_2\rangle\right),\ee
where the subscripts $1,2$ refer to the photon detection times $\tau,2\tau$. Photon 1 has left the loop towards the detectors. Photon 2 remains in the loop, where it is rotated by EPC3 to the $p/m$ polarization basis, resulting in the state $\frac{1}{\sqrt{2}} \left(|h_1p_2\rangle+|v_1m_2\rangle\right)$, which is a two-photon linear cluster in graph representation~\cite{Hein04}.

\begin{figure*}[bt]
\includegraphics[width=\textwidth]{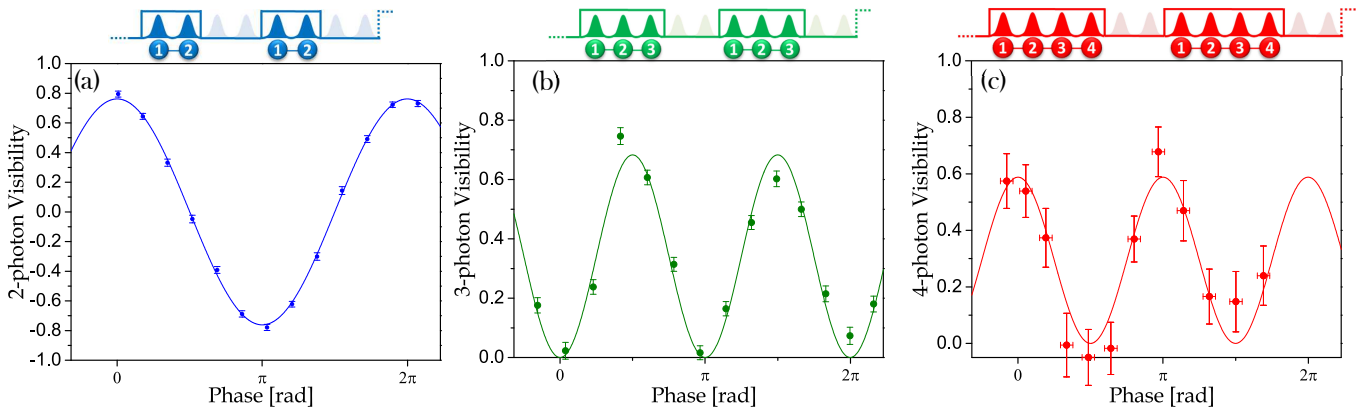}
\caption{\label{Fig3} \textbf{State analysis for 2, 3 and 4 photon cluster state generation} The corresponding pulse sequences are depicted above the graphs. \textbf{(a)} Phase scan for a two-photon state with a visibility of $0.76{\pm}0.01$ at 480\,Hz, 30 seconds per data point. \textbf{(b)} Phase scan for a three-photon state with a visibility of $0.68{\pm}0.03$ at 4.3\,Hz, 11 minutes per data point. \textbf{(c)} Phase scan for a four-photon state with a visibility of $0.59{\pm}0.04$ at 0.04\,Hz, 50 minutes per data point. Visibility error-bars are calculated assuming Poisson distribution. Phase error-bars for (c) are calculated from two-photon residuals fit for multiple scans. Background counts from residual unfiltered pump and detector after-pulsing events are subtracted}.
\end{figure*}

When a third photon enters the setup, it arrives at the entangling PBS1 at the $|p\rangle$ state. The outcome of the entangling PBS1 is
\be\label{GHZ3}
\frac{1}{\sqrt{2}} \left(|h_1p_2\rangle+|v_1m_2\rangle\right)\otimes|p\rangle \rightarrow \frac{1}{\sqrt{2}}\left(|h_1\phi^+_{2,3}\rangle+|v_1\phi^-_{2,3}\rangle\right)\,.
\ee
Thus, the new photon is entangled with the two previous photons into a GHZ state $\frac{1}{\sqrt{2}}\left(|p_1h_2p_3\rangle+|m_1v_2m_3\rangle\right)$, where the photon remaining in the loop (now photon 3) is rotated by EPC3. This three-photon GHZ state is a linear cluster in graph representation.
When a fourth photon enters the setup, repeating the above protocol, the resulting entangled state is not a GHZ state, but the four-photon linear cluster (LC) state
\be\label{Cl4}
\begin{aligned}
|\psi^{(4)}_{LC}\rangle=\frac{1}{2}\Big(
&|p_1h_2h_3p_4\rangle+|p_1h_2v_3m_4\rangle\\
+&|m_1v_2h_3p_4\rangle-|m_1v_2v_3m_4\rangle\Big)\,.
\end{aligned}
\ee 

Local unitary operations on each photon may transform $|\psi^{(4)}_{LC}\rangle$ to other equivalent graph states~\cite{Hein04}.
When more photons come in a timely manner, they are entangled into an ever-growing linear cluster state~\cite{Pilnyak17}. Figure~\ref{Fig2} depicts the quantum logic circuit implementation corresponding to this protocol.

In our protocol, correlations are detected by photon measurements at consecutive time slots. The polarization analysis procedure is performed by applying $\mathbf{X}$ or $\mathbf{Z}$ Pauli operators by EPC4 and PBS2 to the first $n{-}1$ photons. The last $n^{th}$ photon inside the loop is projected on PBS1 (the entangling PBS). If projected onto the $|h\rangle$ polarization, it exits the loop, and if projected onto $|v\rangle$, it stays inside for another cycle. Thus, the last photon polarization is analyzed by its arrival time to either detector.

The projection of $n$ photons results in $2^n$ possible amplitudes. In order to demonstrate the quantum nonlocality of the produced states, the $n$-photon amplitudes are interfered~\cite{Kwiat90,Ou90}.
This nonlocal interference is achieved by rotating the measurement basis of the first $n{-}1$ photons to the diagonal basis via a Hadamard rotation. In this case, half of the photons' probability amplitudes interfere constructively while the other half interfere destructively. The measured amplitude probabilities are used to calculate the visibility $V_n=\mathrm{Tr}\left(\mathbf{X}^{\otimes n}\hat{\rho}\right)$, where $\hat{\rho}$ is the density matrix of the generated $n$-photon cluster state. In order to accumulate more information about the nonlocal interference, the phase $\varphi$ applied by the operator $\mathbf{Z}_\varphi^{\otimes (n{-}1)}{\otimes}\mathbf{I}$ to the $n-1$ photons, where $\mathbf{I}$ is the unit operator and $\mathbf{Z}_\varphi=\mathbf{I}\cos(\varphi/2)-i\mathbf{Z}\sin(\varphi/2)$.
When this phase is scanned, the different amplitude probabilities oscillate, revealing more information about the degree of entanglement (see Supplementary Information).

Figure~\ref{Fig3} shows the resulting phase-dependent oscillations in $V_n$, for the cases of two, three and four photons. Assuming that the main source for imperfect visibilities is the two-photon indistinguishability $M$, the predicted two-, and three-photon visibilities are $V_2(\varphi){=}Mcos(\varphi)$, and $
V_3(\varphi){=}M^2\frac{1{-}cos(2\varphi)}{2}$. The corresponding four-photon visibility $V_4$ has an upper theoretical limit of $\frac{2}{3\sqrt{3}}\simeq0.38$, which limits  the  experimental  sensitivity. Therefore, we present the visibility $V_{4'}{=}\mathrm{Tr}\left(\mathbf{X}{\otimes}\mathbf{I}{\otimes}\mathbf{X}{\otimes}\mathbf{X}\hat{\rho}\right)$ which can reach 1, as this observable is part of the stabilizer group of $|\psi_{LC}^{(4)}\rangle$ before the rotation of the last photon. The dependence of this four-photon visibility on $\varphi$ is $V_{4'}(\varphi){=}M^2\frac{1{+}cos(2\varphi)}{2}$ (see Supplementary Information). Figure~\ref{Fig3}\mred{c} shows our experimental results for $V_{4'}$. The measurements presented in Fig.~\ref{Fig3} were obtained with our brightest sources, corresponding to a negatively charged QD with an in-fiber brightness of about 15\%, and $M{=}0.77{\pm}0.01$.

\begin{figure*}[tbh!]
\includegraphics[width=\textwidth]{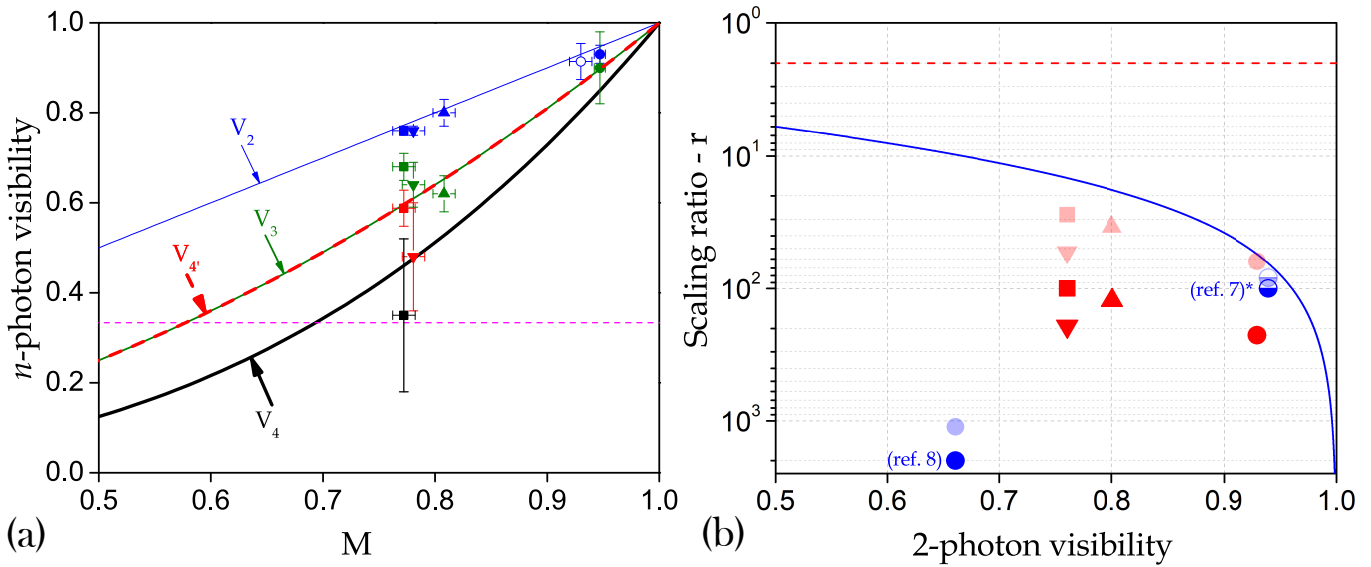}
\caption{\label{Fig4} {\textbf{Entanglement criteria} (a) $n$-photon visibility and indistinguishability $M$ values of various experiments. Solid squares and triangles present results using a charged excitation and solid circles are for the same excitation with spectral filtering. Empty circles are for neutral excitation without spectral filtering. Blue, green, red and black represent $V_2$, $V_3$, $V_{4'}$ and $V_4$ results, respectively. Curves are calculated theoretical values for noise-induced distinguishability. Lines are thicker for increasing $n$. Results for $V_4$ are normalized to one. Horizontal dashed magenta line is the entanglement length visibility threshold. (b) A comparison between scaling ratios as a function of the two-photon visibility for various single-photon source implementations. Light-color data are calculated values for $\eta_d=0.9$. (*) The scaling ratio for~\cite{Zhong18} was calculated based on published photon rates as an heralding single-photon source. Dashed red (solid blue) line represents the theoretical probabilistic gate (heralded PDC sources) limit.} See text for further details.}
\end{figure*}

We have repeated the measurements using several sources based on negatively charged, positively charged or neutral dots, the latter two showing higher M. For some measurements, the indistinguishability was further increased using an etalon filter that removes the remaining phonon sideband emission, but at the cost of a reduced count rate. In Fig.~\ref{Fig4}\mred{a}, the visibility values of these measurements are presented for various values of $M$, obtained by standard Hong-Ou-Mandel (HOM) interference~\cite{HOM87,Santori02}, and the $g^{(2)}(0)$ value from second-order intensity correlations~\cite{HBT56} (see Supplementary Information). The solid lines represent the theoretical expected $n$-photon visibilities, showing good agreement with experiment.

Imperfect photon indistinguishability $M$ limits the entanglement length, defined as the longest possible linear cluster state such that positive concurrence is found between the first and last photons of the chain when all the others are measured~\cite{Pilnyak17}. This length is an upper limit for how far can quantum information flow along the linear state during the one-way quantum computation procedure~\cite{Raussendorf01}. We find that for an $n$-photon entanglement length, the threshold visibility is $V_n{=}\frac{1}{3}$, independent of $n$. All the presented results are above this threshold.
Moreover, our results correspond to maximum entanglement lengths between 23 and 5 photons (see Supplementary Information). In addition to this criterion, the three-photon genuine entanglement can be evaluated by the $V_3$ measure, as the three-photon cluster state matches a GHZ state. The visibility threshold for a three-photon GHZ state is $\frac{1}{2}$~\cite{Mermin90}, which is well exceeded by all our measurements, with values ranging from $V_3{=}0.62{\pm}0.04$ to $V_3{=}0.90{\pm}0.08$, depending on the source used (see Fig.~\ref{Fig4}\mred{a}).

It is instructive to quantify and compare the scalability prospects of various demonstrations. To this end, we define the \textit{scaling ratio} $r$---the reduction factor of detection rates when one photon is added to the protocol. The smaller the scaling ratio, the better the scalability, where the ultimate goal is to reach $r{=}1$, allowing for the deterministic entangling of any number of photons. The detection rate for $n$-photon events is
\be
R_n=R\left(\eta_d\eta_s\eta_l\eta_b\right)^n\eta_g^{n-1},
\ee
where $R$ is the single-photon repetition rate, and the $\eta$'s represent various system efficiencies. Most efficiencies apply to every photon, such as the detector efficiency $\eta_d$, the system loss without the delay loop $\eta_s$, the delay loop (memory cycle) loss $\eta_l$, and the source brightness $\eta_b$, including both its quantum yield and overall optical collection efficiency into a single-mode fiber. One other efficiency doesn't apply to the first photon, the entangling gate efficiency $\eta_g$. Thus, the scaling ratio is
\be
r=R_n/R_{n+1}=\left(\eta_{d}\eta_{s}\eta_{b}\eta_{g}\eta_{l}\right)^{-1}.
\ee

Figure~\ref{Fig4}\mred{b} presents this scaling ratio as a function of the two-photon visibility for various theoretical and experimental situations. 
The use of a probabilistic gate with $\eta_{g}{=}0.5$ limits the scaling ratio to $r{\ge2}$ (dashed red line) considering that $\eta_{d}{=}\eta_{s}{=}\eta_{b}{=}\eta_{l}{=}1$. An intrinsic limitation arises when operating the present scheme with heralded PDC sources. For such sources, the two-photon visibility reduces when increasing the source efficiency. 
The solid blue line represents the dependence of the scaling ratio on the two-photon visibility $r{=}2\frac{1{+}V_2}{1{-}V_2}$, considering $\eta_d{=}\eta_s{=}\eta_l{=}1$ and $\eta_g{=}0.5$ (see Supplementary Information). It represents an  intrinsic upper limit for PDC sources, a limit that could only be overcome by multiplexing schemes \cite{Kaneda19}, yet at the cost of increasingly demanding resources and reduced single-photon repetition rate $R$.

The symbols in Fig.~\ref{Fig4}\mred{b} present scaling ratio values for various implementations. The solid blue symbols correspond to an equivalent scheme but implemented with a PDC source and a free space setup~\cite{Pilnyak17}. The half-filled blue symbols correspond to the predicted data where the same experimental scheme would be implemented using the best PDC source currently available~\cite{Zhong18}. The red symbols correspond to the present work, with measured efficiency values of $\eta_{d}{=}0.25$, $\eta_{s}{=}0.7$, $\eta_{l}{=}0.75$, $0.04{\leq}\eta_{b}{\leq}0.15$ depending on the source used. The light-blue and light-red data points extrapolate previous implementations, predicted ones and the present experimental results to the case $\eta_d{=}0.9$, since such detection efficiencies are currently available at all considered wavelegnths by replacing the silicon APDs with superconducting nano-wire single-photon detectors~\cite{Marsili13}. This allows for a better comparison between different implementations.

The presented comparison shows that our current results, obtained with a lossy and imperfect setup, already reach the upper limit of lossless PDC sources. Further improved scaling ratios are thus expected to be within reach with QD sources in the near future. The QD source brightness can be increased by a factor of up to 2 by changing the excitation schemes following recent propositions~\cite{Wang19,Huber19}. The fibered brightness could also be increased by using a larger numerical aperture collection lens and engineering a better mode matching with the single mode fiber. Moreover, the setup efficiencies $\eta_l$ and $\eta_s$ could also be improved by reducing losses arising mostly from imperfect fiber coupled PBSs. 

In conclusion, we have reported the generation of multi-photon linear cluster states using a single quantum emitter coupled to a compact entangling-loop configuration. The measurement of non-local interference visibilities demonstrates genuine multi-particle entanglement up to four photons. As our protocol relies only on a single quantum emitter and a single entangling gate, the scheme can provide the best possible scalability ratios using linear optics. The present experimental demonstration, although using both imperfect quantum dot sources and entangling apparatus, already demonstrates a scaling ratio on a par with the best possible level predicted for heralded PDC single-photon sources. Straightforward technical improvements, both on the source operation side and on the setup design, will allow reaching larger photon numbers in the near future. Additional delay loops could be used to generate cluster states of higher dimensionality. Finally, removing the last bottleneck of the 50\% probabilistic entangling operation is also foreseeable, considering recent progress in engineering of photon-photon interactions using natural or artificial atoms~\cite{Shomroni14,Tiecke14, Reiserer14,Desantis17}. Thus, the present multi-photon entanglement scheme promises a path for scaling up quantum computation and communication protocols  \cite{Humphreys13, Motes14, Rohde15}.

The authors thank the Israeli Science Foundation for supporting
this work under grants 793/13 and 2085/18. This work was also supported by the ERC PoC PhoW, the QuantERA ERA-NET Cofund in Quantum Technologies, project HIPHOP, the French RENATECH network, a public grant overseen by the French National Research Agency (ANR) as part of the "Investissements d'Avenir" programme (Labex NanoSaclay, reference: ANR-10-LABX-0035). J.C.L. and C.A. acknowledge support from Marie Skłodowska-Curie Individual Fellowships SMUPHOS and SQUAPH, respectively. H. O. Acknowledges support from the Paris Ile-de-France R\'egion in the framework of DIM SIRTEQ. M.E. acknowledges funding by the Deutsche Forschungsgemeinschaft (DFG) (project no. 401390650).

\newpage

\setcounter{equation}{0}
\setcounter{figure}{0}

\renewcommand{\theequation}{S\arabic{equation}}
\renewcommand{\thefigure}{S\arabic{figure}}

\section{Supplementary Information}

\section{Experimental preparation and measurement procedure}
The cluster generation setup uses four electrically driven polarization controllers (EPCs), allowing the transformation of any given input polarization to any desired output polarization of a single photon. Their method of operation uses a local magnetically-induced transverse mechanical stress on the fiber. Each such EPC integrates four channels, where for each channel, the orientation of the applied stress is rotated by an angle of 45 degrees, relative to the neighboring channel. This local change to the fiber refractive index applies any birefringent phase in the range up to $2\pi$, inducing a polarization rotation. Depending on the polarization of the incoming photon, a single channel can be used as a retardation element whereby it can serve as a phase inducing mechanism between different photons of orthogonal polarizations.

The key factor in deciding the fiber loop length, equivalent to a temporal separation of 74~ns, is overcoming the nominal dead-time of the single-photon detectors ($\sim$60~ns). This fiber loop length is equivalent to 6 consecutive laser pulses, under a standard repetition rate of $\sim$81~MHz. Thus, in this cluster generation scheme, each photon generated at the $i^{th}$ time slot was interfered with a photon generated 74~ns later, generated at the $i^{th}{+}6$ time slot.

For an $n$-photon experiment, a train of laser pulses separated by 12.3~ns, used for the excitation of the emitter, is converted by an electro-optical modulator (EOM) into a 74~ns time-separated sequence of $n$ pulses. For example, in the generation of a two-photon cluster state, the excitation sequence consists of an \textit{on-on-off-off} EOM modulation pattern, or equivalently denoted in binary form as 1100. For each \textit{on/off} bin, the EOM transmits/blocks the laser pulses for a duration of $\sim$74~ns. The \textit{off} periods are necessary to reduce the probability of a remaining photon inside the loop before the start of the next sequence. Therefore, the whole temporal duration of the sequence is 296~ns, where each \textit{on} period is comprised of six laser pulses. In this case, a total of 12 photons are injected into the fiber setup, allowing the fiber setup to interfere photons 1-7, 2-8, etc. For the three- (four-) photon experiment, a 11100 (111100) sequence is applied with the EOM, interfering photons in time slots 1-7-13, 2-8-14, etc. (1-7-13-19, 2-8-14-20, etc.).

A comprehensive custom Field-Programmable Gate Array (FPGA) digital electronics hardware was developed for the entire control over the generation of electrical signals for the EOM and data acquisition from the single photon detectors. This system was also implemented with the facility to time delay all electrical signals (input and output) as well as the synchronized collection of detection events into specific time-bin counters. All this was integrated directly into the FPGA fabric. In addition, a custom developed computer software was programmed to collect, analyze and real-time display all the experimental data, as well as control high-current digital to analog converters (DAC) for the correct operation of all EPCs.

The single photon detector signal was collected by the FPGA board into time-bins, thereby assigning time tags to detection events. The $0^{th}$ time-bin denoted the case where a single photon was emitted, did not enter the fiber-loop and immediately triggered a detection event. This is essentially the fastest route by which an emitted photon may be detected. Detection events at any other time-bin, assuming a single photon was emitted, mean that the photon was delayed inside the loop. In this case, the time-bin's index denotes the number of loop iterations experienced by the photon.

\section{Phase channel alignment}
A natural approach to demonstrate the  nonlocal interference would be to set the measurement basis of the photons to 
$\frac{1}{\sqrt{2}} \left(|h\rangle \pm e^{i\theta_i}|v\rangle \right)$ 
and change the delay time between pulses. Yet, it would require changing the repetition rate of the laser. Instead, we set the repetition rate to perfectly match the length of the delay loop, and control a birefringent phase $\varphi$ between the different populations of an $n$-photon entangled state. This phase is added before the polarization rotation inside the delay loop. The polarization controllers EPC1, EPC3 and EPC4 (see Fig.~\ref{Fig1} of the main text) perform the rotations for $\theta_1,\theta_2$ and $\theta_3$, respectively. For the two-photon case, after the PBS projection and post-selection, the following state is produced:
\be\label{phitheta}
|\phi\rangle=\frac{1}{\sqrt{2}} \left(| h_1h_2\rangle+e^{i(\theta_1+\theta_2)}| v_1v_2\rangle \right),
\ee
The addition of a phase between the states results in:
\be\label{phiphi}
|\phi\rangle=\frac{1}{\sqrt{2}} \left(| h_1h_2\rangle+e^{i(\theta_1+\theta_2+\varphi)}| v_1v_2\rangle \right).
\ee
The next step consists on rotating the polarization of the photon exiting the loop into the basis $\frac{1}{\sqrt{2}} \left(| h\rangle \pm e^{i\theta_3}| v\rangle \right)$, and rotating the photon remaining in the loop into the basis $\frac{1}{\sqrt{2}} \left(| h\rangle \pm e^{i\theta_2}| v\rangle \right)$. A visibility measurement in this case would exhibit a dependence on $\varphi$. 

In order to create a controllable phase channel in the fiber system, we align EPC2 and EPC3, where channel 1 in EPC3 (EPC$3_1$) acts as the phase channel of $\varphi$
\be \label{phasechannel}
\begin{aligned}
\mathbf{Z}_{\varphi}&= \left[ \begin{array}{cc}
e^{-i\varphi/2} & 0 \\
0 & e^{i\varphi/2} 
\end{array}\right]
\end{aligned}
\ee

\section{Visibility observable from the stabilizer group}
Cluster states can be defined by the appropriate set of stabilizer group~\cite{Raussendorf03}. These groups are defined by combinations of the Pauli operators $\mathbf{I},\mathbf{X},\mathbf{Y},\mathbf{Z}$. Each stabilizer group comprises the eigenbasis for a specific cluster state.

In our experiment, we describe the produced states before the transformation performed by EPC3 and EPC4. In case of using fast active elements with this scheme, any unitary operation could be performed on any photon. Here, the operation on photons $1$ to $n-1$ is the same while the operation on the $n^{th}$ photon is $\mathbf{X}$, as it is the configuration for the entangling gate of all photons. This imposes the requirement for the observable to be used.

For the two-qubit cluster state 
\be\label{phi}
|\phi^{+}\rangle=\frac{1}{\sqrt{2}} \left(| h_1h_2\rangle+| v_1v_2\rangle \right),
\ee
the corresponding stabilizers are $\mathbf{I}{\otimes}\mathbf{I}$, $\mathbf{X}{\otimes}\mathbf{X}$, $\mathbf{Z}{\otimes}\mathbf{Z}$, and ${-}\mathbf{Y}{\otimes}\mathbf{Y}$.
Thus, for a measurement of $|\phi^{+}\rangle$ with the stabilizer $\mathbf{X}{\otimes}\mathbf{X}$ we expect an eigenvalue 1. The outcome measurement, as changing the state with respect to the phase $\varphi$, is detailed in the next section.

For the three-qubit state 
\be\label{psithree}
|\psi\rangle=\frac{1}{\sqrt{2}} \Big[
|(h+v)_1h_2h_3\rangle 
+|(h- v)_1v_2v_3\rangle\Big],
\ee
$\mathbf{X}{\otimes}\mathbf{X}{\otimes}\mathbf{X}$ is not a part of the stabilizer group and the projection measurement would return zero.
However, applying the phase $\varphi$ described by the operator $\mathbf{Z}_\varphi^{\otimes 2}{\otimes}\mathbf{I}$, setting $\varphi{=}\frac{\pi}{2}$ we get the state
\be\label{psi3i}
|\psi\rangle=\frac{1}{\sqrt{2}} \Big[
|(h+iv)_1h_2h_3\rangle 
+i|(h-iv)_1v_2v_3\rangle\Big],
\ee
for which $\mathbf{X}{\otimes}\mathbf{X}{\otimes}\mathbf{X}$ is part of the stabilizer group and thus the eigenvalue is 1.

The four-photon cluster state produced in this experiment
\be\label{phicl4}
\begin{aligned}
|\psi^{4}\rangle=\frac{1}{2^{\frac{3}{2}}}\Big[
&|(h+v)_1h_2h_3h_4\rangle 
+|(h+v)_1h_2v_3v_4\rangle \\
+ & |(h- v)_1v_2h_3h_4\rangle
-|(h- v)_1v_2v_3v_4\rangle \Big],
\end{aligned}
\ee
is the eigenvector of the stabilizer group with the generators (up to an Hadamard unitary transformation on the last photon these are the generators of the four-qubit linear cluster state) $g_1{=}\mathbf{X}{\otimes}\mathbf{Z}{\otimes}\mathbf{I}{\otimes}\mathbf{I}$, $g_2{=}\mathbf{Z}{\otimes}\mathbf{X}{\otimes}\mathbf{Z}{\otimes}\mathbf{I}$, $g_3{=}\mathbf{I}{\otimes}\mathbf{Z}{\otimes}\mathbf{X}{\otimes}\mathbf{X}$, $g_4{=}\mathbf{I}{\otimes}\mathbf{I}{\otimes}\mathbf{Z}{\otimes}\mathbf{Z}$.

We identify that operating the parity operator $\mathbf{X}^{\otimes4}$ would not give an eigenvalue of 1 as it is not a stabilizer of the state. However, we can identify the operator $g1{\cdot} g3{=}\mathbf{X}{\otimes}\mathbf{I}{\otimes}\mathbf{X}{\otimes}\mathbf{X}$ as part of this group ensuring $|\psi^{4}\rangle$ is an eigenvector. This operator could be measured in the experimental setup without fast active elements. In the next section we derive the behaviour of this when $|\psi^{4}\rangle$ is changed with $\varphi$.

As our scheme enlarges the state in the form of a cluster state with added photons, we can derive a similar observable, as for $n=4$, to any even number of photons. This is deduced from the form of the stabilizer group. We look at the generators of an $n$-qubit cluster (up to an Hadamard unitary transformation on the last photon). The first generators are $g_1{=}\mathbf{X}{\otimes}\mathbf{Z}{\otimes}\mathbf{I}{\otimes}\mathbf{I}{\otimes}\mathbf{I}_{n-4}$, $g_2{=}\mathbf{Z}{\otimes}\mathbf{X}{\otimes}\mathbf{Z}{\otimes}\mathbf{I}{\otimes}\mathbf{I}_{n-4}$, $g_3{=}\mathbf{I}{\otimes}\mathbf{Z}{\otimes}\mathbf{X}{\otimes}\mathbf{Z}{\otimes}\mathbf{I}_{n-4}$, for $2{<}k{<}n{-}1$ the generator would be $g_k{=}\mathbf{I}_{k-2}{\otimes}\mathbf{Z}{\otimes}\mathbf{X}{\otimes}\mathbf{Z}{\otimes}\mathbf{I}_{n-k-1}$, while the last two generators are $g_{n-1}{=}\mathbf{I}_{n-4}{\otimes}\mathbf{I}{\otimes}\mathbf{Z}{\otimes}\mathbf{X}{\otimes}\mathbf{X},g_{n}{=}\mathbf{I}_{n-4}{\otimes}\mathbf{I}{\otimes}\mathbf{I}{\otimes}\mathbf{Z}{\otimes}\mathbf{Z}$, where $\mathbf{I}_{m}{=}\mathbf{I}^{{\otimes} m}$. Multiplication of generators gives rise to the stabilizer group. From certain multiplications between the generators, we find the stabilizer
\be\label{SVnt}
S_{V_{n'}}=\prod_{i=1,3,..}^{n-1} g_i = (\mathbf{X}\otimes\mathbf{I})^{\otimes \left(\frac{n}{2}-1\right)}\otimes\mathbf{X}\otimes\mathbf{X},
\ee
for which the n-qubit state is an eigenvector.

\section{Visibility calculation}\label{VisCalc}
The presented visibilities are calculated from the appropriate projection of the $n$-photon state as described in the main text. Each projection results in $2^n$ populations terms. The general state

\be\label{general_psi}
|\psi^{n}_{\varphi}\rangle=\sum_{k=1}^{2^n} a_{\psi_k}(\varphi)|\psi_k\rangle,
\ee
where $|\psi_k\rangle$ are the $2^n$ states that span the $n$-qubit space in the $h/v$ basis.

For the two-photon state 
\be\label{psi2}
|\psi\rangle=\frac{1}{\sqrt{2}} \left(| h_1h_2\rangle+e^{i\varphi}| v_1v_2\rangle \right).
\ee
measuring along the $\mathbf{X}{\otimes}\mathbf{X}$ basis gives 4 amplitudes
\be\label{2photonamp}
\begin{aligned}
a_{hh}&=a_{vv}=\frac{1}{2^{\frac{3}{2}}}(1+e^{i\varphi})\\
a_{hv}&=a_{vh}=\frac{1}{2^{\frac{3}{2}}}(1-e^{i\varphi})
\end{aligned}
\ee
From this the two-photon visibility is calculated
\be\label{VfromP}
V_2=P_{hh}-P_{hv}-P_{vh}+P_{vv}=cos\left(\varphi\right),
\ee
where $P_{\psi_k}{=}|a_{\psi_k}|^2$.

In a similar way the three-photon state 
\be\label{psi3}
|\psi\rangle=\frac{1}{\sqrt{2}} \Big[
|(h+e^{i\varphi}v)_1h_2h_3\rangle 
+e^{i\varphi}|(h-e^{i\varphi} v)_1v_2v_3\rangle\Big].
\ee
Measuring along $\mathbf{X}^{\otimes 3}$ gives 8 amplitudes
\be\label{3photonamp}
\begin{aligned}
a_{hhh}&=a_{hvv}=\frac{1}{4}(1+2e^{i\varphi}-e^{i2\varphi})\\
a_{hhv}&=a_{hvh}=a_{vhh}=a_{vvv}=\frac{1}{4}(1+e^{i2\varphi})\\
a_{vhv}&=a_{vvh}=\frac{1}{4}(1-2e^{i\varphi}-e^{i2\varphi}).
\end{aligned}
\ee
From this the three-photon visibility is calculated
\be\label{V3fromP}
\begin{aligned}
V_3&=P_{hhh}-P_{hhv}-P_{hvh}+P_{hvv}\\
&-P_{vhh}+P_{vhv}+P_{vvh}-P_{vvv}\\
&=\frac{1-cos(2\varphi)}{2}=sin^2\left(\varphi\right).
\end{aligned}
\ee

\begin{figure}[tb]
\includegraphics[width=\columnwidth]{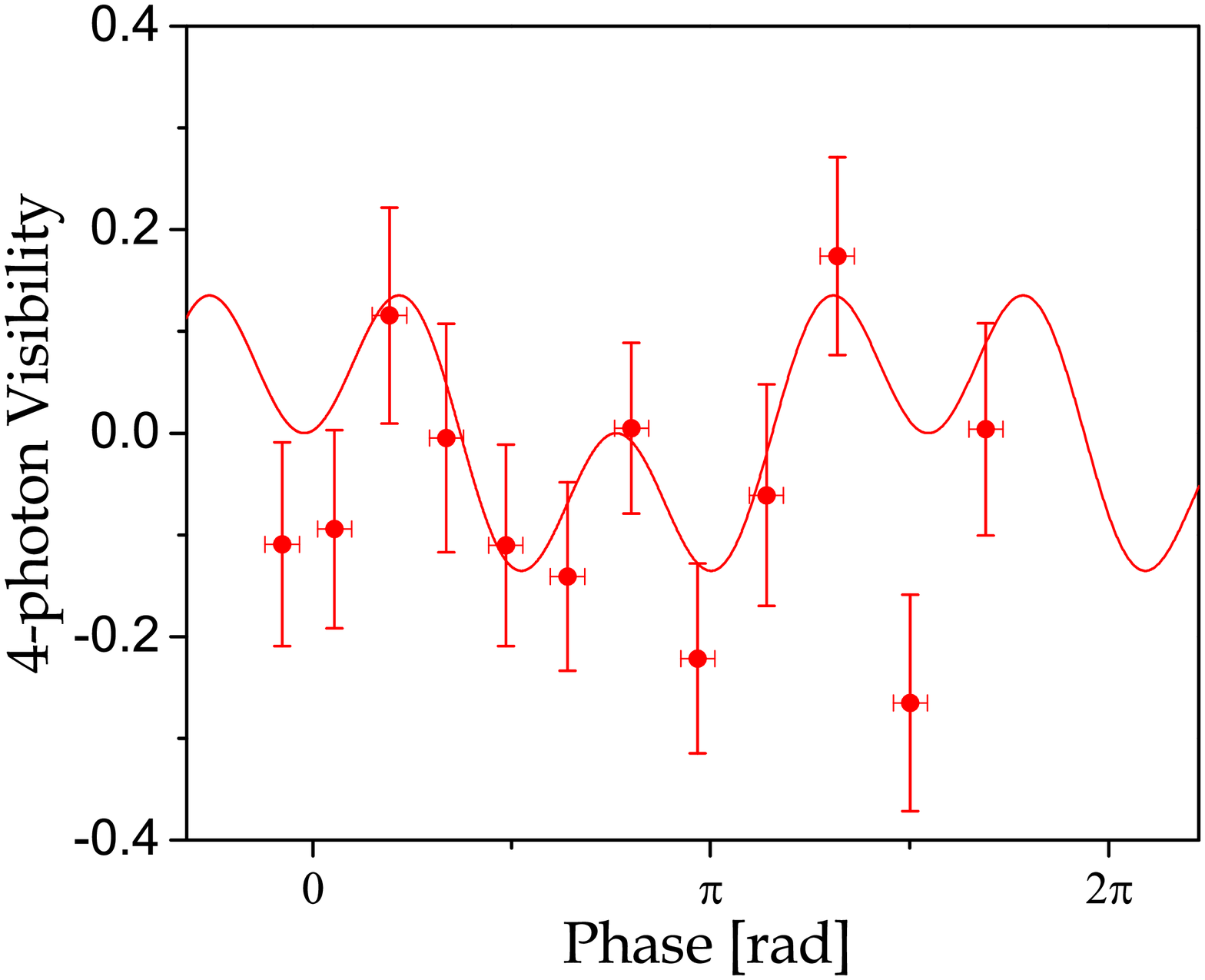}
\caption{\label{FigS1} The normalized visibility $V_4=0.35\pm0.17$.
}
\end{figure}

The four-photon state is
\be\label{phi4}
\begin{aligned}
|\psi^{4}_{\varphi}\rangle=\frac{1}{2^{\frac{3}{2}}}\Big[
&|(h+e^{i\varphi}v)_1h_2h_3h_4\rangle 
+e^{i\varphi}|(h+e^{i\varphi} v)_1h_2v_3v_4\rangle \\
+e^{i\varphi} & |(h-e^{i\varphi} v)_1v_2h_3h_4\rangle
-e^{i\varphi}|(h-e^{i\varphi} v)_1v_2v_3v_4\rangle \Big].
\end{aligned}
\ee
Measuring along $\mathbf{X}^{\otimes 4}$ gives 16 amplitudes
\be\label{4photonamp}
\begin{aligned}
a_{hhhh}&=a_{hhvv}=a_{vhhv}=a_{vhvh}=\frac{1}{2^{\frac{5}{2}}}(1+3e^{i\varphi}-e^{i2\varphi}+e^{i3\varphi})\\
a_{hhhv}&=a_{hhvh}=a_{vhhh}=a_{vhvv}=\frac{1}{2^{\frac{5}{2}}}(1-e^{i\varphi})(1+e^{i\varphi})^2\\
a_{hvhv}&=a_{hvvh}=a_{vvhh}=a_{vvvv}=\frac{1}{2^{\frac{5}{2}}}(1+e^{i\varphi})(1-e^{i\varphi})^2\\
a_{hvhh}&=a_{hvvv}=a_{vvhv}=a_{vvvh}=\frac{1}{2^{\frac{5}{2}}}(1+3e^{i\varphi}+e^{i2\varphi}-e^{i3\varphi})
\end{aligned}
\ee
From this the four-photon visibility, for the observable $S_{V_{4'}}{=}\mathbf{X}{\otimes}\mathbf{I}{\otimes}\mathbf{X}{\otimes}\mathbf{X}$ defined in Eq.~\ref{SVnt} is calculated
\be\begin{aligned}\label{V4fromP}
V_{4'}=&\mathrm{Tr}(S_{V_{4'}}\hat{\rho})=P_1-P_2+P_3-P_4\\
=&\frac{1+cos(2\varphi)}{2}=cos^2\left(\varphi\right),
\end{aligned}
\ee
while for the observable $\mathbf{X}^{\otimes 4}$
\be\begin{aligned}\label{V4X4}
V_4=&\mathrm{Tr}(\mathbf{X}^{\otimes 4}\hat{\rho})=P_1-P_2-P_3+P_4\\
=&\frac{-cos(\varphi)+cos(3\varphi)}{4}=-cos\left(\varphi\right)sin^2\left(\varphi\right),
\end{aligned}
\ee
the maximum this observable can obtain is $\frac{2}{3\sqrt{3}}$.
This visibility from the measured results with a $\pi$ phase is presented in Fig.~\ref{FigS1}.

As the entanglement process of our scheme is a recurring process, in the linear cluster picture it is simply adding another photon and link to the chain, thus we are able to deduce a visibility measure to any $n$-photon state ($n\geq3$)
\be\label{Vn}
V_n=\mathrm{Tr}(\mathbf{X}^{\otimes n}\hat{\rho})=\left(-1\right)^{n-1}cos^{n-3}\left(\varphi\right)sin^2\left(\varphi\right).
\ee
This observable is not part of the stabilizer group of the produced state at $\varphi=0$. In addition, when varying $\varphi$ the maximum amplitude is lower then 1 for $n>3$.
In a similar way we may deduce the $n$-photon visibility to the observable defined in Eq.~\ref{SVnt}. This observable is valid to an even $n$-photon state, as it is part of its stabilizer group, giving unity at $\varphi=0$
\be\label{Vnt}
V_{n'}=\mathrm{Tr}(S_{V_{n'}}\hat{\rho})=cos^{\frac{n}{2}}\left(\varphi\right).
\ee

\section{Visibility and indistinguishability}
\begin{figure*}[tb]
\includegraphics[width=1\textwidth]{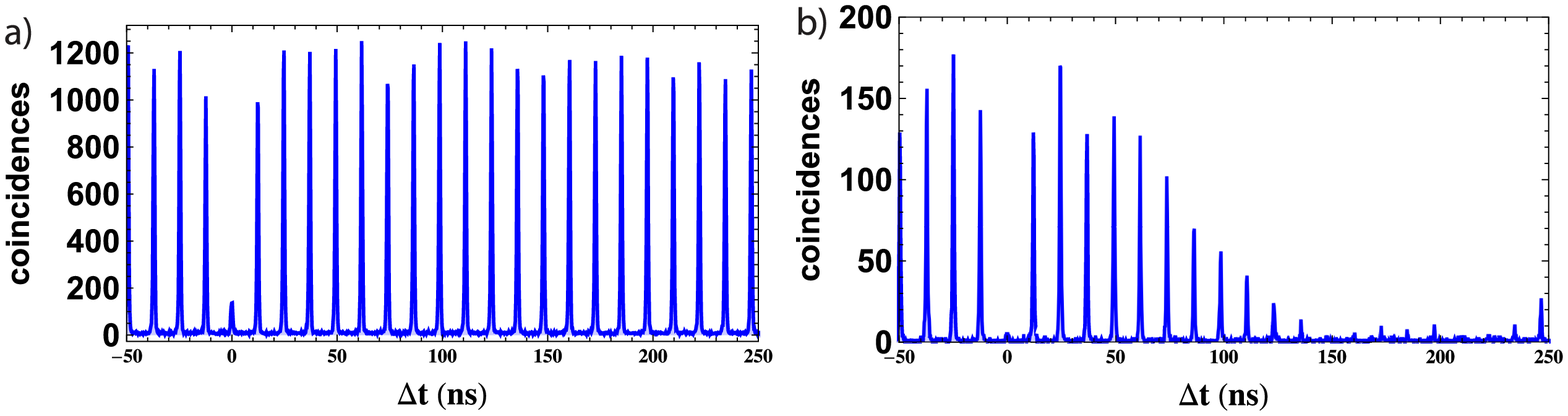}
\caption{\label{FigS2} \textbf{Example of two-photon indistinguishability measurement.} For photons delayed by 12~ns; the measure in panel (a)/(b) is performed under non-modulation/modulation of the train of laser pulses using charged-exciton sources without/with spectral filtering of the single-photon emission.
}
\end{figure*}

A requirement for perfect two-photon visibility $V_2=1$ is obtained when the two interfering photons are indistinguishable $M=1$, and only one photon arrives at each pulse $g^{(2)}(0)=0$. We relate these two terms in a similar manner to the derivation in \cite{Pilnyak17} (appendix A.1). As was previously derived in~\cite{Sun09}, the indistinguishability arises from the mean-wave packet overlap
\be
M=\mathrm{Tr}\left(\hat{\rho}_{i}\hat{\rho}_{j}\right)=\iint_{-\infty}^{+\infty}d\omega_{i}d\omega_{j}f\left(\omega_{i}\right)g\left(\omega_{j}\right)\left|\left\langle \omega_{i}|\omega_{j}\right\rangle \right|^{2}.
\ee
In an Hong-Ou-Mandel (HOM) experiment~\cite{HOM87} the coincidence probability ($P_{cc}$) dip depends on the indistinguishability of two photons from the QD source, when assuming a perfectly balanced beam-splitter (BS) and $g^{(2)}(0)=0$ 
\be
P_{cc}\left(\delta\tau\right)=\frac{1}{2}\left(1-M\left(\delta\tau\right)\right),
\ee
and the measured dip visibility is defined as $V_{HOM}{=}1{-}\frac{P_{cc}(0)}{P_{cc}(\infty)}{=}M$, where $P_{cc}(\infty)$ is the two-photon probability when the single-photons do not arrive simultaneously to the beam-splitter.
When assuming there is a probability of two-photon emission from the source $g^{(2)}(0)>0$, a \textit{lower bound} to the actual indistinguishability is given by:
\be\label{M(g,V)}
M=V_{HOM}+g^{(2)}(0).
\ee
We relate this to the measured visibility interference $V_2$ by the fiber system entangling PBS1 gate followed by the analyzing apparatus~\cite{Mandel91}. When two photons arrive as described to the entangling PBS1, assuming a perfectly balanced PBS and no polarization error, the outcome state is
\be
\hat{\rho}=M\hat{\rho}_{id}+(1-M)\hat{\rho}_{d},
\ee
where $\hat{\rho}_{id}$ ($\hat{\rho}_{d}$) is the the entangled (decohered) state. When performing a measurement of the visibility for the observable $\mathbf{X}{\otimes}\mathbf{X}$ on both photons, $\hat{\rho}_{id}$ would attribute to the traced matrix diagonal elements only two terms $\hat{\rho}_{id}^{1,1}{=}\hat{\rho}_{id}^{4,4}{=}\frac{1}{2}M$. While $\hat{\rho}_{d}$ would equally contribute all terms due to the lack of coherence $\hat{\rho}_{d}^{i,i}{=}\frac{1}{4}(1{-}M)$. Thus, the resulting outcome of the visibility measurement is
\be\label{V2M}
V_2=2(\frac{1}{2}M+\frac{1}{4}(1-M))-2\frac{1}{4}(1-M)=M.
\ee
Yet, when taking into account the probability of two-photon arriving in the same pulse ($g^{(2)}(0)>0$), additional reduction in the measured visibility is present. In the following, we derive the probability of an unwanted two-photon event. There is a 25\% probability that both photons, coming from the same pulse, enter the loop. Then, there is a 50\% probability that these two photons would not exit the loop at the same time, meaning half the probability of a correct event. Thus, their projected state is $|h_1v_2\rangle$ and the measurement along $\mathbf{X}{\otimes}\mathbf{X}$ would yield an equal contribution of all the measured diagonal terms of the density matrix. In such case, the density matrix reads:
\be
\hat{\rho}=\left(1-\frac{1}{2}g^{(2)}(0)\right)\left(M\hat{\rho}_{id}+(1-M)\hat{\rho}_{d}\right)+\frac{1}{2}g^{(2)}(0)\hat{\rho}_{2ph}.
\ee
Both the indistinguishability and the two photon emission affect the resulting visibility
\be
V_2=\left(1-\frac{1}{2}g^{(2)}(0)\right)M.
\ee

We defined two visibility measures in the previous section, $V_n$ for any $n$ and $V_{n'}$ for even $n$. We generalise now the description of these $M$-dependent measures for higher photon number cluster states.

We first model the $n$-photon cluster creation using the PBS entangling gate. For $n$-photon creation, a two-photon post-selection process of the entangling PBS is repeated on the incoming new photon and the photon remaining in the loop $\hat{\rho}^{(n)}_0{=}\hat{\rho}^{(n-1)}{\otimes}\hat{\rho}_p$
\be\label{rhon}
\epsilon(\hat{\rho}^{(n)})=M\varepsilon_0\hat{\rho}^{(n)}_0\varepsilon_0+(1-M)\frac{1}{2}\left(\varepsilon_0\hat{\rho}^{(n)}_0\varepsilon_0+\varepsilon_1\hat{\rho}^{(n)}_0\varepsilon_1\right),
\ee
where $\varepsilon_0{=}\mathbf{I}^{\otimes (n-2)}{\otimes}\left(\mathbf{I}{\otimes}\mathbf{I}{+}\mathbf{Z}{\otimes}\mathbf{Z}\right)$ and $\varepsilon_1{=}\mathbf{I}^{\otimes (n-2)}{\otimes}\left(\mathbf{I}{\otimes}\mathbf{Z}{+}\mathbf{Z}{\otimes}\mathbf{I}\right)$.

Reordering this process, we may regard two contributions which are dependent on $M$. For the two-photon case $\varepsilon_0\hat{\rho}^{(2)}_0\varepsilon_0$ accounts for $\frac{1{+}M}{2}$, while $\varepsilon_1\hat{\rho}^{(2)}_0\varepsilon_1$ accounts for $\frac{1{-}M}{2}$. As resulted in Eq.~\ref{V2M} the visibility measurement $V_2{=}\frac{1{+}M}{2}{-}\frac{1{-}M}{2}{=}M$. 
When performing the $\mathbf{X}^{\otimes 3}$ measurement on $\hat{\rho}^{(3)}$ we are again considering the process two terms that act on $\hat{\rho}^{(2)}{\otimes}\hat{\rho}_p$ thus $V_3{=}M\left(\frac{1+M}{2}{-}\frac{1-M}{2}\right){=}M^2$. This is recurring to any $n$, $V_n{=}M^{n-2}\left(\frac{1+M}{2}{-}\frac{1-M}{2}\right){=}M^{n-1}$. When considering $V_n'$, we are performing the measurement when $n$ is even. For $n{=}4$ the $\mathbf{X}{\otimes}\mathbf{X}$ part of the measurement would give the $M$ dependence, and the $\mathbf{X}{\otimes}\mathbf{I}$ would give $\frac{1+M}{2}{-}\frac{1-M}{2}{=}M$, so $V_{4'}{=}M^2$.
Thus, the dependence of the maximum amplitude of $V_n$ on the distinguishability is $M^{n-1}$, while $V_{n'}$ scales as $M^{\frac{n}{2}}$.

Several quantum dot sources were used for the present measurements. We show in Fig.~\ref{FigS2} two indistinguishability measurements. The indistinguishability histogram is measured in a path-unbalanced Mach-Zehnder interferometer. One of the arms has an optical length $k{\times} \Delta t$ longer than the other, $k$ being an integer number and $\Delta t$ the temporal distance between consecutive pulses. Thus, allowing to interfere two successively emitted photons on the second fiber beam-splitter of the interferometer. At the output of this fiber beam-splitter we place two single-photon detectors to carry out a time-correlated two-photon coincidence detection. The simultaneous coincidence counts at $\Delta t{=}0$ allow to quantify the degree of photon indistinguishability between the two interfering photons. Figure~\ref{FigS2} demonstrates this for a 12~ns delay. 

It is important to remark that the different pulse sequences used for the generation of different photon cluster state sizes ($n{=}2,3,4$ photons, see Fig.~\ref{Fig3} of the main text) involve trains of emitted single photons separated by 74~ns. Therefore, the photon indistinguishability involved in the generation of cluster states is slightly lower than the one extracted in Fig.~\ref{FigS2}, due to QD spectral wandering~\cite{Loredo16}.

Panel (a) corresponds to a negatively charged exciton which showed higher brightness. Such a source with was used for the four-photon experiment. 
Without any spectral filtering, the measured $M$ amounts to $M{=}0.78{\pm}0.01$. Note that this indistinguishability is substantially lower than the one observed on both neutral and positively charge excitons, indicating  nuclear spin induced dephasing. The measurement shown in panel (b) was performed on positively charged exciton using a 10 pm spectral filtering to reduce the phonon sideband contribution resulting in an indistinguishability of $M{=}0.95{\pm}0.01$. The latter is taken with the pulse-modulation turned on, leading to the triangular envelope shape of the coincidence histogram.

The presence of an electron or a hole in the quantum dot is controlled following the method described in \cite{Hilaire19}. Maximum occupancy is obtained by finely adjusting the power and wavelength of a continuous wave non-resonant laser with wavelength typically between 890 nm and 905 nm.

\section{Background counts}
The background correction was implemented by subtracting background measurement counts $N^{bg}$ from the measured coincidence counts (CC) $N^{meas}$. Phase scans were performed acquiring CC data, $N^{meas}_i$, $i{=}1..2^n$ according to the desired pulse sequence - $\underbrace{1 1...1}_{n} 00$, where $n$ is the number of passing pulses, corresponding to the length of the desired chain. As the phase scan completes, it is repeated in the form of background measurements, $n{-}1$ pulses are open meaning no CC of an $n$-photon state should be recorded. As there are $n$ combinations corresponding to these background measurements $1^{(1)}...0^{(k)}...1^{(n)} 00$, this process is repeated for each combination with $k{=}1..n$, recording the background CC $N_{i,k}^{bg}$. The final outcome $N_i$, which is the full pulse signal CC subtracting the background counts, is thus
\be\label{bg_sub}
N_i=N^{meas}_i-\sum_{k=1}^n N_{i,k}^{bg}.
\ee

The source of the background counts is from photons constantly entering the fiber system with no synchronization to the EOM pulse sequence. This is partly due to the EOM extinction ratio (1:100), which amounts to at most $1\%$ of the coincidences. Here, the majority of the background signal is due to a residual single photon emission arising from the non-resonant laser pump used to insert a charge in the quantum dot. It amounts to typically ${\sim}10\%$ of the coincidences for a coincidence window of 5\,ns. In a future experiment, this technical issue could be addressed in different ways, either by modulating the non-resonant pump laser, or using a smaller coincidence window.

\section{Entanglement length}
\begin{figure}[tb]
\includegraphics[width=\columnwidth]{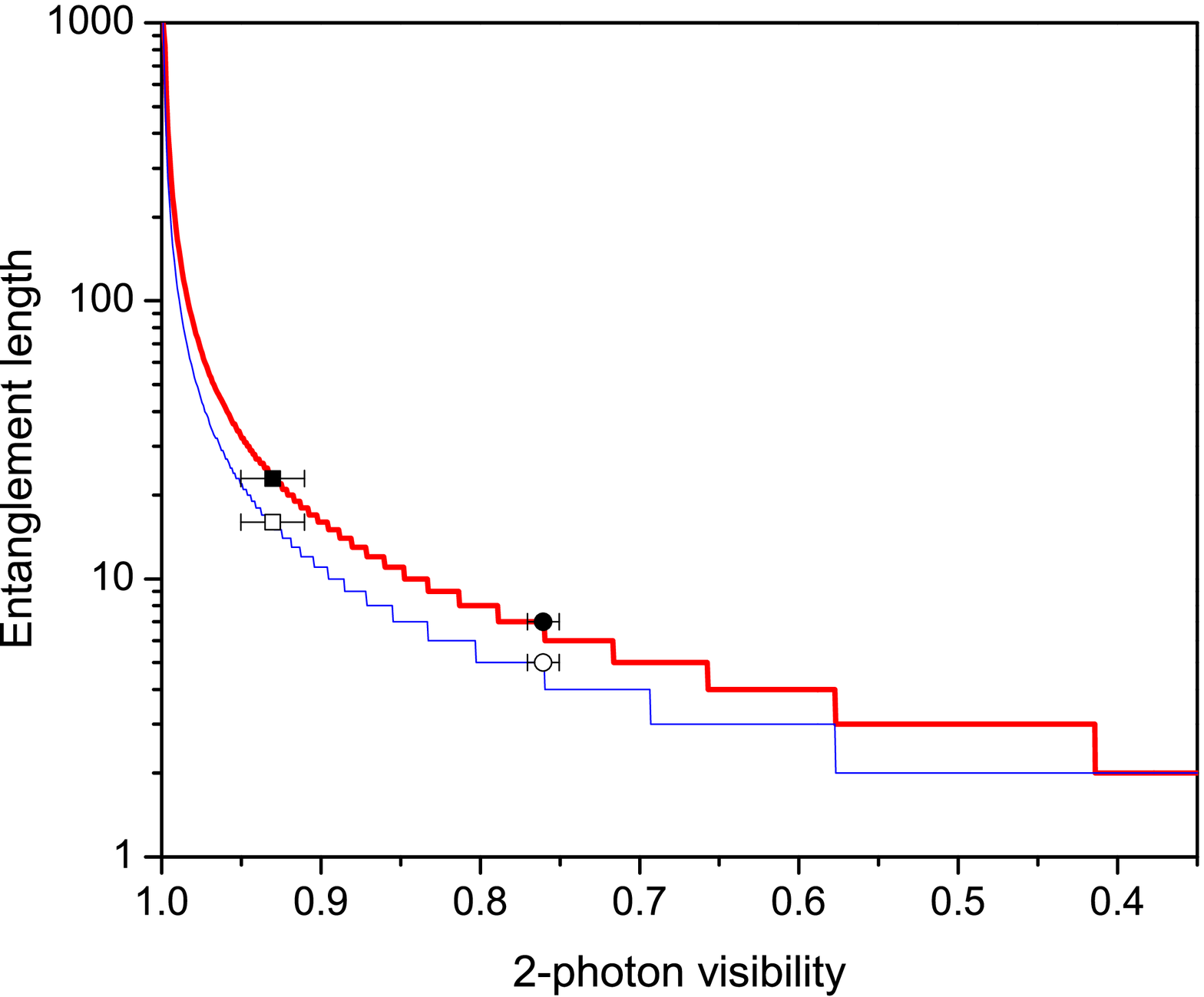}
\caption{\label{FigS3} \textbf{Entanglement length (EL) dependence on the two-photon visibility.} The red thick and blue thin plots represent the calculation for distinguishing and depolarizing noise, respectively. For the measured two-photon visibilities of $93\pm2\%$ and $76\pm1\%$ the corresponding entanglement length is 23 (solid square) and 7 (solid circle) for the distinguishing noise and 16 (open square) and 5 (open circle) for the depolarizing noise. Errors are calculated from fitting to experimental data.
}
\end{figure}
The entanglement length measure $\mathcal{L}$ was introduced and analyzed in \cite{Pilnyak17}. It is defined as the longest possible linear cluster state that will result with a pair of photons with positive concurrence after all photons, but the first and last of the chain, are measured along the $\hat{y}$ axis. This length is an upper limit to how far can quantum information flow along the linear state during the one-way quantum computation procedure~\cite{Raussendorf01}. The main noise factor taken is the indistinguishability of photons, modeled by the number of modes $N_m$. This corresponds to the two-photon visibility in the diagonal basis $N_m{=}\frac{1}{V_2}$. For $n$-photon creation, the process is detailed in Eq.~\ref{rhon}.

In this work we add the possibility of a general "white noise" as the source of noise in the process.

It is of interest to evaluate the presence of genuine multi-party entanglement in the created states, and extrapolate the system potential if time was not a practical constraint. For that purpose we use the entanglement length measure. Figure~\ref{FigS3} presents the dependence of this entanglement length on the measured two-photon visibility, assuming two different noise sources. In our experiment, the lack of perfect indistinguishability between the consequently emitted photons accounts for most of the reduced visibility. Calculating the entanglement length under this noise source results with the red thick line. It accounts for the possibility of a genuinely entangled linear state of 7-23 photons depending on the source $M$ factor. The two-photon visibility can also be inferred directly from the visibilities of states of more photons. The measured four-photon visibility implies a state of 6 genuinely entangled photons as well. Moreover, when considering the worst case scenario of a uniformly depolarizing noise source, it provides a lower limit for the entanglement length for a measured two-photon visibility. In case of such a noise source, the dependence on the photon source $V_2$ factor will result in 5-16 entangled photons. Thus, we have shown that regardless of the applied noise model, the presented four photon linear cluster state is genuinely entangled.

When regarding $V_2$ to originate from some general noise process, in a similar to dependence derived from Eq.~\ref{rhon}, we find it dependence $V_2{=}(1{-}\delta)$, where $\delta$ is the amount of noise introduced - for the case of indistinguishability noise $M{=}1{-}\delta$. Again we can generalize this to any $n$, so $V_n{=}(1{-}\delta)^{n-1}{=}V_2^{n-1}$. 

For a certain linear cluster state, that could be represented by a chain of length $n$, we may set a threshold value to the minimal value of $V_2$ for which the entanglement length is the chain length, $\mathcal{L}_{\min V_2}(n){=}n$. From observing the solid blue thin line in Fig.~\ref{FigS3}, the calculation gives a value of $V_2$ that satisfies the threshold condition. For $n{=}2$ it is clear that $\min V_2{=}\frac{1}{3}$. For any added photon in the chain, this factor increases, as it is the minimal value between two photons before the removal of one from the chain. Thus, for any number of $n$, $\min V_2{=}\left(\frac{1}{3}\right)^\frac{1}{n-1}$. Meaning, the larger the state is, the higher the two-photon visibility should be to sustain the entanglement length.

We can further determine a threshold for the minimal $n$-photon visibility $\min V_n$. From the previous relations we come to $\min V_n {=} \min V_2^{n-1} {=} \frac{1}{3}$ for any $n$.

\section{Parametric down-conversion scaling ratio}
We start with the state definition of an $n$ photon pair in mode $a$ and $b$ produced in the Parametric Down-Conversion (PDC) process
\be
|\psi\rangle=\sqrt{1-|\lambda|^2}\sum_{n=0}^{\infty}|\lambda|^n|n\rangle_a|n\rangle_b,
\ee
this is a two-mode squeezed vacuum with strong quantum correlations~\cite{Kok07}. $\lambda$ is the squeezing parameter with its absolute square proportional to the pump power. For this derivation we assume non photon-number discriminating detectors (`bucket') with 100\% efficiency.
The probability for a single pair emission $|1\rangle_a|1\rangle_b$ is $P^{(1)}(1){=}(1{-}|\lambda|^2)|\lambda|^{2}$. The rate for $N$ consecutive photon pairs emission is then $\left(P^{(1)}(1)\right)^N{=}\left((1{-}|\lambda|^2)|\lambda|^{2}\right)^N$. When using this source as a single photon source in our scheme, each pair contributes one photon to the entangled state (the other photon being the heralded). Yet, as we are not employing photon-number-resolving detectors any $|k\rangle_a|k\rangle_b$, $1\leq k\leq n$ per pulse would be considered as a single pair emission. Thus, the probability for a detection in modes $a$ and $b$ is 
\be
P^{(n)}(1)=1-P(0)=1-(1-|\lambda|^2)=|\lambda|^{2}.
\ee
The rate for $N$ consecutive photon pairs emission is then $R_N{=}\left(P^{(n)}(1)\right)^N{=}\left(|\lambda|^{2}\right)^N$.
The scaling ratio is then the probability of adding another pair 
\be
r=\frac{R_N}{R_{N+1}}=\frac{1}{|\lambda|^{2}}.
\ee
From the description of the interaction Hamiltonian of PDC $\lambda{=}tanh(\tau)$, where the interaction parameter $\tau$ is a linear function of the non-linear crystal properties and the pump electrical field strength, meaning an increase in $r$ reflected in the brightness is achievable by increasing one of the parameters $\tau$ depends on~\cite{Eisenberg04}.

We now turn to examine the two-photon visibility dependence on $\tau$~\cite{Takeoka15,Eisenberg04}. As presented in Fig.~\ref{Fig4}\mred{b} in the main text, we assume only physical considerations, meaning perfect detection efficiency, perfect optical and polarization components and perfect indistinguishability between photons. In the PDC apparatus employing the photon's polarization degree of freedom, the visibility is defined using a polarization analyzing system followed by 4 single photon detectors. Coincidence counts are recorded and the visibility is defined as
\be
V_2=\frac{P_{14}+P_{23}-P_{13}-P_{24}}{P_{14}+P_{23}+P_{13}+P_{24}}.
\ee
From this definition we are left with a relation between the visibility and $\tau$
\be
V_2=\frac{1-tanh^2\tau}{1+tanh^2\tau}.
\ee
We are now able to relate the scaling ratio dependence on the interaction parameter
\be
r=\frac{1+V_2}{1-V_2},
\ee
we may notice that an increase in $r$ would come at a cost of reduced visibility. This relation is displayed as the solid line in Fig.~\ref{Fig4}\mred{b} in the main text.

\end{document}